\documentstyle[12pt]{article}
\def\beq{\begin{equation}}
\def\eeq{\end{equation}}
\def\bea{\begin{eqnarray}}
\def\eea{\end{eqnarray}}
\def\bq{\begin{quote}}
\def\eq{\end{quote}}

\def\gappeq{\mathrel{\rlap {\raise.5ex\hbox{$>$}}
{\lower.5ex\hbox{$\sim$}}}}

\def\lappeq{\mathrel{\rlap{\raise.5ex\hbox{$<$}}
{\lower.5ex\hbox{$\sim$}}}}

\begin{document}
\pagestyle{empty}
\begin{flushright}
{CERN-TH/97-356}\\
{hep-ph/9712340}\\
\end{flushright}
\vspace*{5mm}
\begin{center}
{\bf HUNTING DOWN INTERPRETATIONS OF THE HERA
LARGE-$Q^2$ DATA}
\\
\vspace*{1cm} 
{\bf John ELLIS} \\
\vspace{0.3cm}
Theoretical Physics Division, CERN \\
CH - 1211 Geneva 23 \\
\vspace*{2cm}  
{\bf ABSTRACT} \\ \end{center}
\vspace*{5mm}
\noindent
Possible interpretations of the HERA large-$Q^2$ data are reviewed
briefly. The possibility of statistical fluctuations cannot be ruled out,
and it seems premature to argue that the H1 and ZEUS anomalies are
incompatible. The data cannot be explained away by modifications of parton
distributions, nor do contact interactions help. A leptoquark
interpretation would need a
large $\tau q$ branching ratio. Several $R$-violating
squark interpretations are still viable despite all the constraints,
and offer interesting experimental signatures, but please do not hold
your breath.
\vspace*{1cm}

\begin{center}
{\it Invited Parallel Session Talk}\\
{\it  presented at the}\\
{\it Europhysics  Conference on High-Energy Physics} \\
{\it Jerusalem, August 1997} 
\end{center}
\vspace*{3cm}

\begin{flushleft} CERN-TH/97-356 \\
December 1997
\end{flushleft}








%

\section{Introduction}

The large-$Q^2$, large-$x$ HERA data reported by the H1~\cite{H1} and
ZEUS~\cite{ZEUS} collaborations
early in 1997 excited considerable interest, being subject to many theoretical
interpretations and stimulating many related experimental searches. The new
instalment of HERA data and other experimental data reported here cast new
light on many of the suggested interpretations. Now is an opportune moment to
take stock of the situation, assess which interpretations remain viable, and
review how those remaining might be elucidated by future analyses. Space
restrictions do not permit me to do justice to the many papers on this
subject, and what follows is a personal selection of material: for other
recent reviews, see~\cite{Reviews}.

\section{Statistical Fluctuations?}

H1 and ZEUS reported here the current status of their 1997 neutral-current data
analysis~\cite{Elsen}, in addition to the data published
previously~\cite{H1,ZEUS}. H1 has by now found 7+1=8
events vs. 0.95+0.58 = 1.53 ($\pm$ 0.29) background in the window $M_e = 200
\pm $12.5 GeV, $y_e > 0.4$~\cite{par31302}, and ZEUS has 4+1=5 events vs.
0.91 + 0.61 =
1.51 ($\pm$
0.13) background in the window $x > 0.55, y_{D_A} > 0.25$~\cite{par31301},
and they quote
respectively probabilities of 1\% (1.9\%) for such a fluctuation in any
window (in the quoted window).
 These probabilities are not negligible, but one should be impressed if one
could be convinced that the two experiments were seeing compatible
anomalies --
more on this shortly. It is important to note that, although the 1997 data do
not repeat the magnitudes of the anomalies reported in the earlier data, this
year's data are not below the rates expected in the Standard Model.

In addition, the HERA collaborations have reported apparent excesses in their
charged-current data: H1 has found 14 events vs. $Q^2 > 10^4$ GeV$^2$,
compared
with 8.33 $\pm$ 3.10 expected~\cite{par31302}, and ZEUS has 15 events vs
9.4
$\pm$ 2.5 $\pm$
1.6 expected~\cite{par31301}. These excesses are intriguing, but not as
significant as those
reported in the neutral-current samples.

For the moment, we conclude that statistical fluctuations cannot be
excluded,
even if they are formally unlikely.

\section{Are the H1 and ZEUS Anomalies Compatible?}

The reported excesses are apparently in different kinematic regions: H1
discusses the region $M_e = 200 \pm 12.5$ GeV (corresponding na\"\i vely to
$x_e = 0.45 \pm 0.06$) and $y_e > 0.4$~\cite{par31302}, whereas ZEUS
discusses $x_{D_A} > 0.55$
and $y_{D_A} > 0.25$~\cite{par31302}. However, the two collaborations use
different measurement
methods -- H1 using electron measurements and ZEUS the double-angle method
-- with different statistical and systematic errors. In particular, the
methods
are affected differently by initial-state photon
radiation~\cite{AEGLM,Wolf,Drees}, the possible
effects of gluon radiation have not been evaluated completely, and detector
effects such as the energy calibrations cannot be dismissed entirely.

The most complete analysis, using all the available measured quantities,
indicates that the experiments' apparent mass difference $\Delta m = 26\pm 10$
GeV should be reduced to 17 $\pm$ 7 GeV, on the basis of which it seems that
the decay of a single narrow resonance is
unlikely~\cite{Bassler,par31302}.

In my view, it is premature to exclude a resonance interpretation. I personally
would like to see a more complete treatment of gluon radiation using the full
perturbative-QCD matrix element now available~\cite{PQCD}, the Standard
Model backgrounds
are non-negligible, and one should perform a global fit to all the large-$x$
data of both collaborations. Nevertheless stocks in models with production of a
single leptoquark or $R$-violating squark decaying into $eq$ final
states are surely going down.

\section{Modified Parton Distributions?}

Forgetting for the moment about possible structures in the $x$ distributions,
could the global excess at large $x$ and $Q^2$ be accommodated by modifications
of the conventional parton distributions used to predict Standard Model rates?
One suggestion was a possible ``lump" in the valence quark distribution at
$x\sim 1$ at low $Q^2$, but evolution of this ``lump" does not have a large
effect in the HERA range of $x$ and $Q^2$~\cite{CTEQ}. An alternative
suggestion was
intrinsic charm~\cite{GV}, but this is tightly constrained by EMC data on
$\mu N
\rightarrow \mu\bar ccX$, and unable to modify significantly the Standard Model
predictions if $c(x) = \bar c(x)$~\cite{BKM}. However, even if one makes
$\bar c(x)$
harder than $c(x)$, the effect on the HERA neutral-current rate is $\lappeq$
10\% in the range $M = 200 \pm 10$ GeV, though effects up to 50\% might be 
possible in the charged-current rate, and this type of model might provide an
alternative interpretation of the CDF high-$E_T$ jet excess~\cite{MT}.

My conclusion is that the proposed modifications of the parton distributions
cannot make a significant contribution to explaining the neutral-current data,
but they might make a more important contribution to resolving the possible
charged-current excess. This is in any case less significant, and I would
recommend waiting to see how it develops.

\section{New Contact Interactions?}

Those relevant to the HERA data may be written in the general form
\beq
{\eta\over m^2}(\bar e\gamma_\mu e)_{L,R}(\bar q\gamma^\mu
q)_{L,R}~:~\eta=\pm
1,~ q = u,d
\label{1}
\eeq
giving 16 possibilities, which are tightly constrained by experiments on parity
violation in atoms (which, however, allow parity-conserving combinations:
$\eta^{eq}_{LR} = \eta^{eq}_{RL}$ and other, more general, possibilities), by
LEP2 measurements of $e^+e^-\rightarrow\bar qq$, and by CDF measurements of the
Drell-Yan cross section. A recent global analysis~\cite{Wisc} of
atomic-physics
parity-violation experiments, lower-energy deep-inelastic $eN$ experiments,
LEP, Drell-Yan and the $Q^2$ distribution of the first instalment of
large-$Q^2$ HERA data found fits with $\chi^2$ = 176.4 for 164 degrees of
freedom with no additional contact interactions: $\chi^2 =
187.1$ for 163 d.o.f. with $\eta^{eu}_{LR} =
\eta^{eu}_{RL}$, which is {\it worse}, and $\chi^2$ = 167.2 for 156 d.o.f. when
all possible contact interactions are allowed, which ``represents no real
improvement over the Standard Model". The real killer for
the contact-interaction fits is the constraint imposed by the CDF Drell-Yan
data. Needless to say, no contact interaction would give a resonance peak
in the $eq$ invariant mass! Moreover, the constraints of weak universality
make it difficult to arrange a large charged-current signal. 

Therefore, I personally conclude that there is no
motivation currently to pursue further the possibility of contact interactions.

\section{Leptoquark?}

The tree-level cross section for producing a leptoquark or $R$-violating
squark
in $ep$ collisions is
$\sigma = {(\pi/ 4s)}~~\lambda^2 ~~ F_J$,
where $F_J = 1(2)$ for scalar(vector) leptoquarks. One-loop QCD corrections to
this cross section have been calculated, as well as a number of
final-state
kinematic distributions~\cite{PQCD}. Using the known parton distributions
at $x \sim$ 0.5,
it was found that, in order to explain the first batch of HERA data, one needed
$\lambda\simeq 0.04/\sqrt{B(e^+q)}$ if production was via $e^+d$ collisions, and
$\lambda\gappeq 0.3/\sqrt{B(e^+q)}$
for production off some $q$ or $\bar q$ in the sea, where $B(e^+q)$ is the
branching ratio for decay into the observed $e^+$-jet final
states~\cite{AEGLM}. 

The 
dominant production mechanism at the FNAL Tevatron collider is pair production
via QCD, which is independent of $\lambda$, and for which the one-loop QCD
corrections have also been calculated~\cite{FNALQCD}. Using these,
CDF~\cite{CDF} and D0~\cite{D0} together
imply $m_{LQ} \gappeq$ 240 GeV for $B(e^+q) = 1$ for a scalar leptoquark, and a
far stronger (though somewhat model-dependent) limit for a vector leptoquark,
which option we shall not discuss further. Could the leptoquark have $B(e^+q) <
1$?  A large $B(\bar\nu q)$ is difficult to reconcile with charged-current
universality~\cite{notCCuniv,AGM}. Also, if $B(e^+q) > B(\bar\nu q)$ as
suggested by the possible
relative magnitudes of the neutral- and charged-current anomalies (19
events
with 8 background vs 7 events with 3 background for $M >$ 175 GeV, $Q^2 >$
15,000 GeV$^2$)~\cite{par31302,par31301}, the FNAL data together imply
$m_{2Q} >$ 220 GeV, beyond the
range suggested by the HERA data~\cite{Janot}. Moreover, a significant
$B(\mu q)$ is
excluded by upper limits on anomalous muon capture on nuclei: $\mu N
\rightarrow e N$~\cite{AEGLM}. The only remaining unexcluded decay mode
into
Standard Model particles is $\tau q$, and it would be worthwhile looking for
this decay mode at FNAL and/or HERA, so as to pin down or exclude finally the
leptoquark interpretation. Failing $\tau q$, one needs some decay mode
involving particles beyond the Standard Model, and these are provided by the
final interpretation we discuss.

\section{Squarks with $R$ Violation?}

The supermultiplet content and symmetries of the minimal supersymmetric
extension of the Standard Model (MSSM) admit extra couplings beyond those
responsible for the quark and lepton masses:
$\lambda_{ijk}L_iL_jE^c_k + \lambda^\prime_{ijk}L_iQ_jD^c_k +
\lambda^{\prime\prime}_{ijk}U_i^c D^c_j D^c_k$.
Any one of these couplings would violate $R = (-1)^{3B+L+2S}$. They would
provide dilepton, leptoquark and dijet signatures, respectively, and a
combination of the $\lambda^\prime$ and $\lambda^{\prime\prime}$ couplings
would cause baryon decay. The apparent HERA excess could be due to a
$\lambda^\prime$ coupling:
$\lambda^\prime_{ijk}L_iQ_jD^c_k \Rightarrow \lambda^\prime_{ijk} e^+_R
d_{R_k}
\bar{\tilde u}_{L_j}~,~~
\lambda^\prime_{ijk} e^+_R \bar u_L \tilde d_{R_k}$ 
via the specific mechanisms $e^+d\rightarrow {\tilde u}_L, {\tilde
c}_L, {\tilde t}$ with $\lambda^{\prime} \sim 0.04$, or
$e^+ s (b) \rightarrow {\tilde u}_L, {\tilde c}_L, {\tilde t}$ and
conceivably $e^+ {\bar u} ({\bar c}) \rightarrow \tilde
d, \tilde s, \tilde b$
with $\lambda^\prime\sim 0.3$~\cite{CR,AEGLM,BKMW,DM,KK}. The absence of
neutrinoless $\beta\beta$ decay~\cite{betabeta}
imposes
\beq
\vert\lambda^\prime_{111}\vert < 7\times 10^{-3}~\left({m_{\tilde q}\over
200~{\rm GeV}}\right)^2~~\left({m_{\tilde g}\over 1~{\rm TeV}}\right)^2
\eeq
which rules out $e^+d\rightarrow\tilde u_L$ (and $e^+\bar u\rightarrow \tilde
d)$, but allows $e^+d\rightarrow \tilde c_L, \tilde t$. The upper limit on
$K^+\rightarrow\pi^+\bar\nu\nu$ decay~\cite{Knunu} imposes
\beq
\vert\lambda^\prime_{ijk}\vert\lappeq 0.02 \left({m_{\tilde d_{R_k}}\over
200~{\rm GeV}}\right) 
\eeq
up to mixing and cancellation factors,
which barely allows the $e^+d\rightarrow \tilde c_L$ mechanism. Many 
constraints on large $\lambda^\prime$ couplings, such as charged-current
universality, atomic-physics parity violation and neutrino masses exclude
almost all sea production mechanisms. The only mechanisms that survive this
initial selection are the $e^+d\rightarrow \tilde c_L$ (down-scharm),
$e^+d\rightarrow \tilde t$ (down-stop) and $e^+s\rightarrow \tilde t$
(strange-stop) interpretations~\cite{CR,AEGLM,BKMW,DM,KK}.

The strange-stop interpretation is quite severely constrained by precision
measurements at LEP1, notably the $\rho$ parameter, which requires non-trivial
$\tilde t_L-\tilde t_R$  mixing, and limits on the violation of
universality in
$Z^0\rightarrow e^+e^-, \mu^+\mu^-$ decay, which impose
$\vert\lambda^\prime_{13j}\vert\lappeq 0.6$, so that $B(\tilde
t\rightarrow
e^+q)$ cannot be very small~\cite{ELS}. On the other hand, we recall that
the FNAL upper
limits on leptoquark production require $B(e^+q)\gappeq 0.5$ for $m\sim$ 200
GeV~\cite{Janot}.

Competitive branching ratios for the $R$-violating decay $\tilde c_L\rightarrow
e^+d_R$ and the $R$-conserving decay $\tilde c\rightarrow c_L\chi$ (followed by
$R$-violating decay of the lightest neutralino $\chi$) are possible in generic
domains of parameter space, as a result of a cancellation in the
$R$-conserving coupling of the lightest neutralino~\cite{AEGLM}. This is
much more
difficult to arrange in the down-stop interpretation, where $\tilde
t\rightarrow e^+d$ either dominates (if $m_{\tilde t} < m_t + m_\chi)$ or is
dominated by $\tilde t\rightarrow t\chi$ (if $m_{\tilde t} > m_t + m_\chi)$.
The strange-stop scenario is intermediate: after taking into account the
LEP
constraints, there is very little parameter space where $B(\tilde t\rightarrow
e^+s) \sim B(\tilde t\rightarrow t\chi)$ if $m_{\tilde t}$ = 200 GeV, but
considerably more if $m_{\tilde t}\sim$ 220 GeV~\cite{ELS}.

At the moment none of the three supersymmetric 
interpretations is very healthy, but none
can yet be ruled out.

\section{Tests to Discriminate Between Models} 

More statistics should soon clarify the $Q^2$ and $x$ distributions, whether
there is really an excess, telling us
whether it is peaked in any particular mass bin, and
whether there is any unusual pattern of gluon radiation.

We should also know soon whether there is really an excess of charged-current
events, which would be difficult to understand in most scenarios invoking new
contact interactions, leptoquarks, or even $R$-violating squarks (except
in some
interesting corners of parameter space, where cascade squark decays imitate the
simple charged-current topology~\cite{AGM,CDRW}). In 1998, it is planned
to run HERA in $e^-p$
mode: most scenarios predict the absence of a signal, though one could appear
in the $e^+s\rightarrow\tilde t$ scenario.

Turning to other signatures, in the $e^+d\rightarrow\tilde c_L$ scenario one
predicts observable $\tilde c\rightarrow c\chi$ decays followed by
$\chi\rightarrow\bar qqL, \bar qq\nu$. There should be equal numbers of
${\ell}^\pm$
+ jets final states, and there could be $\mu^\pm$ and/or $\tau^\pm$ + jets as
well as $e^\pm$ + jets. These decay modes should also be observable at the FNAL
Tevatron collider~\cite{AEGLM}. A first search for them has been
made~\cite{CDF}, but it does not yet
have the required sensitivity.

Effects could show up in $e^+e^-\rightarrow \bar qq$ at LEP2, either via
contact interactions or via virtual leptoquark or squark
exchange~\cite{LEPx,par31306}, which might
be detectable in models invoking production from the sea. Identification of the
final-state quark flavours would improve the sensitivity, e.g., to $\tilde t$
exchange in $e^+e^-\rightarrow\bar ss$.

One exciting possibility is the production of a direct-channel $\tilde\nu$
resonance if there is a $\lambda_{ijk}L_iL_jE^c_k$ coupling~\cite{LEPdir}.
The LEP
experiments have already provided some limits on $\vert\lambda\vert$ as a
function of $m_{\tilde\nu}$~\cite{L3}. Unfortunately, sensitivity is
rapidly lost if
$E_{cm} < m_{\tilde\nu}$. It would be a shame if LEP missed out on
discovering
a squark or  a sneutrino because it was not pushed to the maximum possible
centre-of-mass energy, so let us all hope that a way can be found to operate
LEP at its design energy of 200 GeV, with an integrated luminosity
sufficient to exploit fully its capabilities.



%

\end{document}